\documentclass[]{interact}

\usepackage{epstopdf}
\usepackage[caption=false]{subfig}

\usepackage[numbers,sort&compress]{natbib}
\bibpunct[, ]{[}{]}{,}{n}{,}{,}
\renewcommand\bibfont{\fontsize{10}{12}\selectfont}
\makeatletter
\def\NAT@def@citea{\def\@citea{\NAT@separator}}
\makeatother

\theoremstyle{plain}

\theoremstyle{definition}

\theoremstyle{remark}

\usepackage{url}
\graphicspath{{images/}}
\usepackage{multirow}
\usepackage{multicol}
\usepackage{tabularx}

\begin{document}

\articletype{RESEARCH PAPER}

\title{SuperMat: Construction of a linked annotated dataset from superconductors-related publications}

\author{
\name{Luca Foppiano\textsuperscript{a}\thanks{Corresponding authors: FOPPIANO.Luca@nims.go.jp, ISHII.Masashi@nims.go.jp}, Sae Dieb\textsuperscript{a}, Akira Suzuki\textsuperscript{a}, Pedro Baptista de Castro\textsuperscript{b}, Suguru Iwasaki\textsuperscript{b}, Azusa Uzuki\textsuperscript{b}, Miren Garbine Esparza Echevarria\textsuperscript{b}, Yan Meng\textsuperscript{b}, Kensei Terashima\textsuperscript{b}, Laurent Romary\textsuperscript{c}, Yoshihiko Takano\textsuperscript{b}, Masashi Ishii\textsuperscript{a}}
\affil{\textsuperscript{a}Material Database Group, MaDIS, NIMS, Tsukuba, 305-0044, Japan; 
\textsuperscript{b}Nano Frontier Superconducting Materials Group, MANA, NIMS, Tsukuba, 305-0047, Japan;
\textsuperscript{c}ALMAnaCH, Inria, Paris, 75012, France}
}

\maketitle








\begin{abstract}
A growing number of papers are published in the area of superconducting materials science.
However, novel text and data mining (TDM) processes are still needed to efficiently access and exploit this accumulated knowledge, paving the way towards data-driven materials design. 
Herein, we present SuperMat (Superconductor Materials), an annotated corpus of linked data derived from scientific publications on superconductors, which comprises 142 articles, 16052 entities, and 1398 links that are characterised into six categories: the names, classes, and properties of materials; links to their respective superconducting critical temperature (\textit{T\textsubscript{c}}); and parametric conditions such as applied pressure or measurement methods.
The construction of SuperMat resulted from a fruitful collaboration between computer scientists and material scientists, and its high quality is ensured through validation by domain experts. 
The quality of the annotation guidelines was ensured by satisfactory Inter Annotator Agreement (IAA) between the annotators and the domain experts. 
SuperMat includes the dataset, annotation guidelines, and annotation support tools that use automatic suggestions to help minimise human errors.
\end{abstract}


\begin{quote}
    \textbf{CLASSIFICATION}: Databases, data structure, ontology
\end{quote}

\section{Introduction}
The vast majority of scientific knowledge exists as published articles~\cite{Grigas2017JustGI, Khabsa2014TheNO, OrduaMalea2015MethodsFE, Bjrk2009ScientificJP}. 
These publications are presented mainly as text, which is challenging to be used as a machine-readable structure. 
Meanwhile, as a part of the text and data mining (TDM) discipline, computer-assisted information collection from the literature has become a supportive asset for scientific research~\cite{doi:10.1063/5.0021106}. 
In the past decades, new TDM processes were developed for several natural science disciplines to achieve automatic document processing such as information retrieval, entity extraction, and clustering.  
TDM has been applied in biology for identifying interactions between agents (e.g. bacteria, viruses, genes, and proteins)~\cite{10.1371/journal.pone.0004554, Krallinger2010, Krallinger2009ExtractionOH} to support the research on serious diseases including cancer~\cite{Krasnitz2019CancerB}. 
In chemistry, it was used for the disambiguation of chemical compounds names, synthesis extraction, and retrieval~\cite{Hawizy2011ChemicalTaggerAT}.
In both domains, the application of TDM was based on manually curated datasets (corpora) that functioned as infrastructures. Examples are the BioCreative IV CHEMDNER corpus~\cite{Krallinger2015TheCC} in chemistry, and Genia~\cite{Kim2003GENIAC} and GENETAG~\cite{Tanabe2005GENETAGAT, Ohta2009IncorporatingGA} in biology. Such datasets are crucial for developing, training, and evaluating TDM systems.

In comparison, such resources in the materials science domain are rather limited. 
Reported cases include NaDev~\cite{Dieb2016} on nanocrystal devices research, a corpus for extracting synthesis recipes~\cite{kononova_text-mined_2019}, and ChemDataExtractor~\cite{court2018auto} which focuses only on chemical entities. In the superconductors domain, we could identify MagDb~\cite{court_magnetic_2020} focusing on magnetic materials with limited information categories. Another project is SC-CoMIcs~\cite{yamaguchi-etal-2020-sc}.
SuperMat is different from SC-CoMIcs based on the following reasons: (a) it provides full papers instead of abstracts which contain more detailed information about the research on superconducting materials, and (b) it contains linked entities. 

To address this shortage of infrastructure, experimental data is extracted manually~\cite{doi:10.1021/cm400893e}, or ab-initio calculations are used~\cite{Jain2013CommentaryTM_materialsProject} but they might not accurately describe the real system.
Several challenges still hinder the data-driven exploration of materials (also called Materials Informatics (MI)), namely: the lack of data standard, infant stage of the data-driven culture, a wide variety of conflicting stakeholders, and missing incentives for researchers to contribute to large collaborative initiatives~\cite{Hill2016MaterialsSW}. 
To bridge these gaps, it is necessary to create infrastructural resources to support TDM processes in materials science through the automatic construction of databases for materials and their properties. 
Such application can minimise the need for humans to read the new papers and extract the key information therein. 
Equally importantly, it enables scientists to focus and leverage computing power and human resources to find deeper relationships between superficially unrelated information. 
Other applications include providing semantically enriched search engines that accept fine-grain queries~\cite{Liu2019SurfaceMR} to reduce the time needed to access specific information. 
These processes cannot be established without essential resources such as dictionaries, lexicons, and datasets. 

Research on superconducting materials has been growing rapidly towards both fundamental science as well as practical applications. Superconductors display many intriguing phenomena including zero-resistivity, the ability to host a high magnetic field, quantisation of the magnetic flux, and vortex pinning.  
Current applications of superconductors include medical instruments, high-speed trains, quantum computers, and the Linear Hadron Collider (LHC)~\cite{PhilippeBook, Kizu2010ConstructionOT, Cardani2017NewAO}. 
However, discovering a new superconductor is a challenging task~\cite{PhysRevB.103.014509}. For example, in a previous work~\cite{doi:10.1088/1468-6996/16/3/033503} out of $\sim$1000 studied materials, only 3\% were found to be superconducting. 
The National Institute for Materials Science (NIMS) in Japan has been manually constructing databases to support material research, and SuperCon (\url{http://supercon.nims.go.jp}) is a manually curated data source for the superconductor domain.
These databases would help researchers design new superconducting materials with a higher superconducting critical temperature (\textit{T\textsubscript{c}}) (ideally up to room temperature)~\cite{Hamlin2019SuperconductivityNR,stanev2017machine}.
However, the current resources are very limited and not dynamic enough to incorporate the information from new publications in a timely manner. 
In this paper, we present SuperMat (Superconductors Materials), an annotated linked corpus for superconducting material information. 
This dataset contains 142 documents with 16052 (7166 unique) entities, and 1398 links that can serve as an infrastructural data for TDM processes in the domain of superconducting materials. 
We also describe the construction guidelines for SuperMat, in the hope of supporting researchers to systematically create annotated data.
Furthermore, the unique feature of links between entities in SuperMat will allow the development of more precise methodologies to associate a particular material with its properties.

\label{sec:method}
\section{Methods}

\label{content-acquisition}
\subsection{Content acquisition}
SuperMat originates from PDF documents of scientific articles related to superconductor research. 
The PDF format is the most widely used format for scientific publications \cite{johnson2018pdfStatistics}.
The original documents were collected from the following sources: (a) the Open Access (OA) version of peer-reviewed articles referenced in the SuperCon database records; 
(b) articles provided by domain experts containing suitable items and potential links of material names, \textit{T\textsubscript{c}} values, measurement methods, and pressures; (c) articles from "condensed matter" category of arXiv (\url{https://arxiv.org/archive/cond-mat}) selected using the search terms of "superconductor", "critical temperature", and "superconductivity". 

Pre-print versions of peer-reviewed articles were obtained using a lookup service for bibliographic data called biblio-glutton (\url{https://github.com/kermitt2/biblio-glutton}) that aggregates data from various sources: the Crossref (\url{https://www.crossref.org/}) bibliographic database, the unPaywall (\url{http://unpaywall.org}) service, the PubMed Central repository (\url{https://pubmed.ncbi.nlm.nih.gov/}), and mappings to other databases. 
We queried \textit{biblio-glutton} using the bibliographic data of each article referenced in Supercon; subsequently, we downloaded the pre-print article associated with the retrieved record, if available. 
Although the published version may be different from the pre-print version of a document, the differences measured by comparing pre-print and peer-reviewed articles in biology~\cite{carneiro_comparing_2020} measured objective differences to be around 5\%.

\subsection{Preliminary annotation study}
\label{subsec:preliminary-annotation-study}
Preliminary annotation study was carried out to assess the effort required from the annotators to reach an acceptable Inter Annotation Agreement (IAA \textgreater 0.7) .
We annotated two randomly selected OA papers, by using a preliminary version of the guidelines with a limited tag-set of four labels: \texttt{<material>}, \texttt{<tc>} (expression describing the presence or absence of superconductivity), \texttt{<tcValue>} (value of \textit{T\textsubscript{c}}), and \texttt{<doping>} (amount of substitution, such as stochiometric values, usually expressed as functions of x or y).
The process was iterated multiple times.
Each iteration ended with computing the IAA using the Krippendorff's alpha coefficient~\cite{Krippendorff2004ReliabilityIC,Zapf2016MeasuringIR}, while annotators discussed the disagreements, and updated the guidelines.

Based on the results in Table~\ref{table:summary-preliminary-annotation}, IAA reached a satisfactory level around 0.9 after the third iteration. 
In the second iteration, although the average IAA reached 0.7 on three of the four labels, the average agreement was not satisfactory. 
When analysing the disagreement, we noticed that the low score in the \texttt{<doping>} label was caused by a heavy overlap with the \texttt{<material>} label, which required more precise definition in the guidelines. 

Based on this preliminary study, the following changes were implemented. 
(a) The label \texttt{<doping>} was merged under the \texttt{<material>} because, even with detailed documentation it was too difficult for humans to annotate them in a consistent way.
(b) Three more labels were added: measurement methods and pressure (described as parametric conditions in relation to \textit{T\textsubscript{c}}), and class of materials. 

\subsection{Tag-set design}
\begin{figure}[ht]
  \centering
  \includegraphics[width=\linewidth]{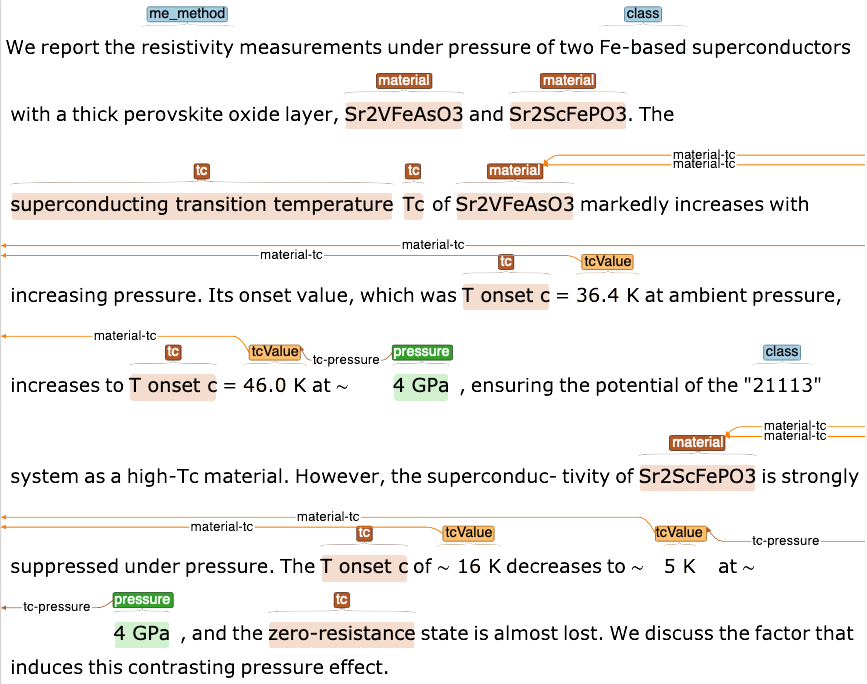}
  \caption{Example in the annotated corpus. The excerpt was taken from~\cite{Kotegawa2009ContrastingPE}.}
  \label{fig:example-annotations-and-links}
\end{figure}

The tag set (also referred to as \textit{labels}) represents the classes of entities and the type of links between them, which were designed to be extracted from the text (Figure~\ref{fig:example-annotations-and-links}).

\subsubsection{Entities}
Entities (also referred as Named Entities, mentions, or surface forms) are chunks of texts that represent an information of interest, as follow: 

\begin{itemize}
\item Class (tag: \texttt{<class>}) represents a group of materials defined by certain characteristics.
Superconducting materials can be classified according to different criteria such as the composition and magnetic properties. 
Among publications collected for this study, the domain experts identified three types of classes based on: (a) the composition and crystal structure, (b) material phenomena (e.g. "I-type" and "II-type superconductivity", "BCS superconductors", "nematic", and "conventional/unconventional superconductivity"), and (c) high/low \textit{T\textsubscript{c}} value (e.g. "high-tc” superconductors). 

In this work, we only considered the (a) classes, mainly because the material composition and crystal structure do not change with time. For example, a cuprate from 1998 is still called a cuprate today. 
In comparison, many material phenomena used for (b) are not robust enough, and can be biased by the viewpoint of the author(s) or research group, or the measurement methods. 
Finally, the definition of "high-tc" superconductors (c) is completely relative; i.e., with the progress of research, materials once considered "high-tc" might not be so anymore.

\item Material (tag: \texttt{<material>}) identifies the name of one or more materials. 
This label is used to collect the following types of information: 
\begin{itemize}
    \item Chemical formula indicating the material by its general or stochiometric formula (e.g. \texttt{LaFe\textsubscript{1-x}O\textsubscript{7}}, \texttt{WB\textsubscript{2}}),
    \item Compositional name (e.g. \texttt{magnesium diboride}) or abbreviations (e.g. \texttt{YBCO}), 
    \item The material's shape (e.g. wire, powder, thin film) or form of material (e.g. single/poly crystal), 
    \item Modification by a dopant (\texttt{Zn-doped}, \texttt{Si-doped}) or by percentage of doping (\texttt{2\%-doped}). We also considered qualitative expressions such as \textit{overdoped}, \textit{lightly doped}, and \textit{pure} as valid information, 
    \item Substrate information (e.g. \texttt{grown on MgO(100) film}) when it was adjacent to the material name or formula, in the text,
    \item Additional information about the sample (e.g. \texttt{as-grown}, \texttt{untwinned}, \texttt{single-layer}) when it was adjacent to the material name or formula, in the text. 
\end{itemize}

\item Superconducting critical temperature (tag: \texttt{<tc>}) identifies expressions related to the phenomenon of superconductivity. Any temperature mentioned in the text is not necessarily the \textit{T\textsubscript{c}}. Rather, it could refer to the temperature for other processes/events such as annealing/sintering temperature, specific measurements, and structural changes.
This label identifies the presence or absence of superconductivity at a given temperature.
In addition, modifiers of this information (increasing/descreasing \textit{T\textsubscript{c}}) are also retained. 

\item Superconducting critical temperature value (tag: \texttt{<tcValue>}) represents the temperature at which the superconducting phenomenon occurs. 
It can be defined by different experimental criteria, such as the onset, mid-point of resistivity drop, or zero resistivity.
This value also considers boundary conditions, such as the \textit{onset of superconductivity}, \textit{zero resistance}. 

\item Applied pressure (tag: \texttt{<pressure>}) indicates the applied pressure corresponding to a measured \textit{T\textsubscript{c}}. 

\item The measurement method (tag: \texttt{<me\_method>}) indicates the method used to measure or calculate the presence of superconductivity. Here, we considered the following categories: resistivity, magnetic susceptibility, specific heat, and theoretical calculations. 
\end{itemize}

\subsubsection{Links}
The links connects entities of materials or samples to their corresponding properties, conditions, and results. 
The links are non-directional, and there are no restrictions on the number of links for each entity. 
We defined three types of links:

\begin{itemize}
    \item material-tc: linking materials to their \textit{T\textsubscript{c}} values.
    \item tc-pressure: connecting \textit{T\textsubscript{c}} and the applied pressure under which it was obtained.
    \item tc-me\_method: linking \textit{T\textsubscript{c}} and the corresponding measurement method. 
\end{itemize}

\subsection{Annotation guidelines}
\label{subsec:annotation-guidelines}
Annotation guidelines include the principles and the rules that describe  what constitutes as desired information for the SuperMat dataset and how to annotate it. They include detailed description of the specific rules that have been defined for each type of information to be annotated, with one or more definitions and examples illustrating what to annotate in different cases, exceptions, and references. We used an online system to track the discussions and decisions when a question or a comment was raised, and provided a link to such issues in the respective description or example. 
In addition, the guidelines include \textit{linking rules} that provide information on how to correctly connect the entities in a relationship. 
The guidelines were built using a dynamic markup language (called RestructuredText) and stored in a git (\url{https://git-scm.com/}) version control system repository. We deployed them as HTML files via web, which were updated automatically after each modification. They can be accessed at \url{https://supermat.readthedocs.io}.

\subsection{Annotation support tools}
\label{subsec:annotation-support-tool}
The task of annotating documents is tedious and requires both attention and subject knowledge from the annotators.
Annotation support tools aim to maximise the efficiency of annotators and minimise human mistakes. 
They are composed of a web-based collaborative annotation tool, automatic annotation suggestions, and automatic corpus analysis. 

\subsubsection{Web-based collaborative annotation tool: INCEpTION}
\label{subsec:annotation-tool}

\begin{figure}[htb]
    \centering
    \includegraphics[width=\linewidth]{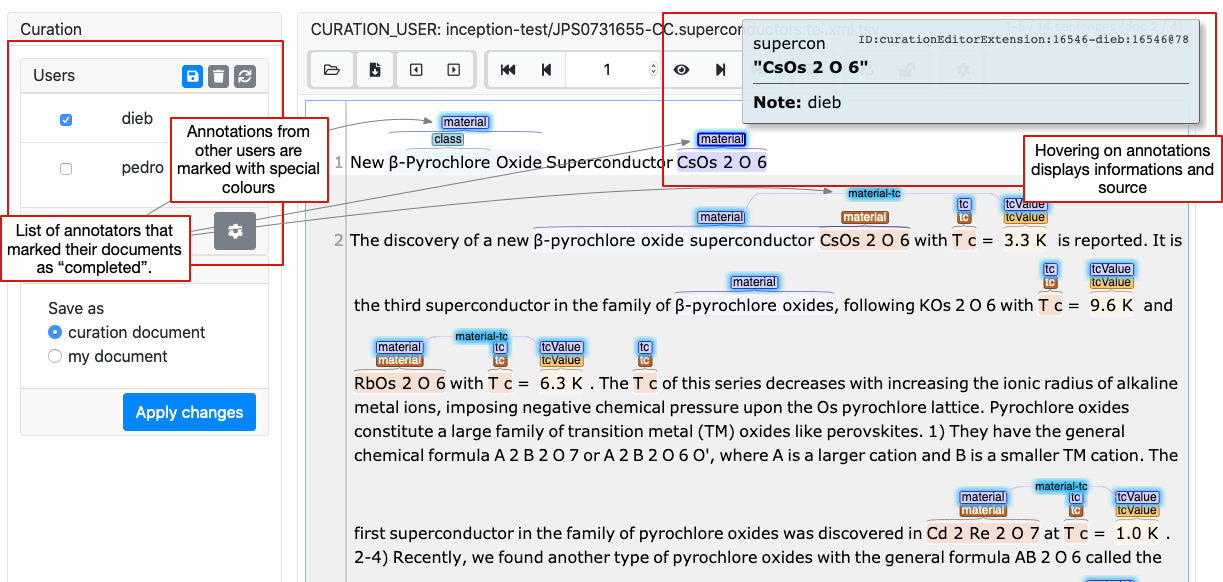}
    \caption{INCEpTION curation interface. The example is taken from~\cite{Yonezawa2004NewO}.}
    \label{fig:inception-curation-interface}
\end{figure}

The annotation tool is the platform used for creating, correcting and linking annotations.
After evaluating several tools, we selected INCEpTION~\cite{tubiblio106270,eckart-de-castilho-etal-2016-web}, a web-based multi-user platform for machine-assisted rapid dataset annotation construction. 
INCEpTION provides supportive functionalities that include: 
\begin{itemize}
    \item Multi-layer annotation sheets allow different annotation schemas over the same documents, 
    \item Two annotation steps: annotation consists of manually correcting pre-imported documents, while curation allows another user to validate the annotations (Figure~\ref{fig:inception-curation-interface}). 
    \item On-the-fly automatic suggestions based on active learning and string matching (Figure~\ref{fig:inception-curation-interface}), 
    \item Bulk annotation corrections, and 
    \item Being open-source (Apache 2.0 license), and under active development at the time of this paper (\url{https://inception-project.github.io/}).
\end{itemize}

\subsubsection{Annotation suggestions}
\label{subsec:automatic-system-prototype}

Previous works have demonstrated that annotation suggestions improve the quality of the output~\cite{Fort2010InfluenceOP,Nvol2011SemiautomaticSA,Lingren2014EvaluatingTI}.
We provide two types of annotations suggestions. 
(i) \textit{Machine-based annotated data} that were assigned to the documents before loading into the annotation tool. Here, we use a machine learning (ML)-based system from a previously implemented prototype~\cite{foppiano2019proposal} to support our tag-set. 
(ii). \textit{Active learning recommendations} provided by INCEpTION are assigned on-the-fly based on previous annotations. 
The active-learning recommendations are less precise since they aim to increase the recall, and therefore they need to be explicitly accepted by the annotator.

\subsubsection{Automatic corpus analysis}
Automatic corpus analysis is a set of scripts designed to run after the validation step. 
These scripts automatically find inconsistencies in the links and entities, while extracting the statistics of the corpus. 
We calculated the inconsistencies by examining every annotated entity and computing the frequency of the same text being annotated with different labels. 
The script outputs a summary table by visualising each annotation value, as well as their labels and frequencies.
We visually inspected this table, because the reported inconsistencies can be either obvious mistakes (Table~\ref{table:dataset-inconsistencies-clear}) or arise from ambiguities (Table~\ref{table:dataset-inconsistencies-unclear}); therefore their context should be verified. 

Although the links are conceptually non-directed, we have defined a practical convention to maintain their consistency. For example, \textit{material-tc} is always represented as a link between \texttt{<tcValue>} and \texttt{<material>} entities. 
The script also computes the statistics (Table \ref{table:summary-content}) for the number of entities (total, unique, by class), the number of links (total, intra- and inter-paragraph, between paragraphs), and other statistical information.

\subsection{Annotation process}
\label{subsec:annotation-workflow}
\begin{figure}[htb]
\centering
  \centering
  \includegraphics[width=0.5\linewidth]{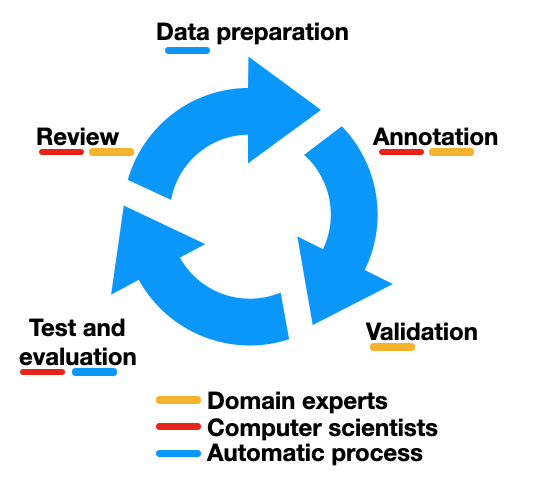}
  \caption{Annotation workflow. Different colours illustrate the involvement of each group at each step of the workflow.}
  \label{fig:schema-comparison-modified-workflow}
\end{figure}

The annotation workflow (Figure~\ref{fig:schema-comparison-modified-workflow}) was designed following the \textit{MATTER} (Model, Annotate, Train, Test, Evaluate, and Revise) schema\cite{pustejovsky2012natural} and other related work~\cite{Dieb2016, Krallinger2015TheCC}.
The workflow is composed of five steps (Figure~\ref{fig:schema-comparison-modified-workflow}): \textit{data-preparation}, \textit{correction}, \textit{validation}, \textit{testing and evaluation}, \textit{revision}. 
This workflow involves three main actors: the automatic process, computer scientists, and the domain experts.

The first step of the annotation process involves preparing the machine-based annotated data from the source PDF documents. 
The PDF files are converted to an XML-based format, and annotation is automatically applied. 
This is followed by four more steps: 

\begin{itemize}
\item Annotation: The human annotator can select a document and manually add, remove, or modify each entity based on rules defined in the guidelines. Once the annotation is complete, the document is marked "ready" for the validation. 

\item Validation/Curation by domain experts: Annotations from different users are validated and merged into a final document (Figure~\ref{fig:inception-curation-interface}). 
The domain expert ("curator"), can compare the different annotated versions, and select the best combination of annotations, or add new ones. 
This step ensures that the annotations are cross-checked and that the document is validated by domain experts.

\item Automatic consistency checks and statistical analysis: This step aims to discover obvious mistakes such as mislabelling or incorrect linking. 
A sequence labelling model is trained and evaluated using 10-fold cross-validation. The evaluation provides precision, recall, and f-score metrics for all the labels.
The resulting model is used for producing machine-based annotated data in the following iteration.

\item Review: Retrospective analysis of the past iteration, where unclear cases are discussed and documented in the annotation guidelines. 

\end{itemize}

\subsection{Data transformation}
\label{subsec:transformation-of-data}
\begin{figure}[htb]
    \centering
    \includegraphics[width=\linewidth]{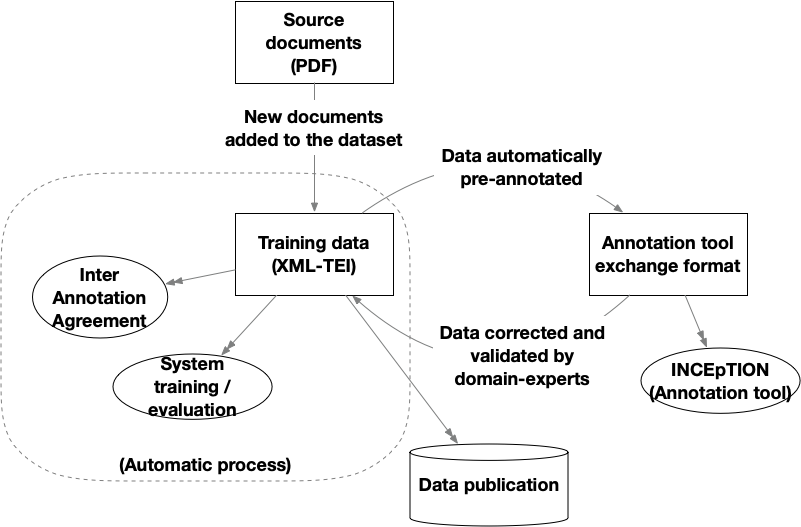}
    \caption{Summary of the data transformation flows.}
    \label{fig:data-transformation}
\end{figure}

There are two processes of data transformation (Figure~\ref{fig:data-transformation}): (a) from the source document (PDF) to the dataset format representation (XML-based), and (b) from the dataset format representation to the annotation tool exchange formats (\url{https://inception-project.github.io/releases/0.16.1/docs/user-guide.html\#sect_formats}) and vice-versa. 
\begin{itemize}
    \item PDF to XML-based: This step converts the PDF source document to the dataset format representation in XML following the Text Encoding Initiative (TEI, \url{https://tei-c.org/}) format guidelines. 
    Such transformation is performed by leveraging the functionalities provided by GROBID (\url{https://github.com/kermitt2/grobid}).
    
    We developed a customised process for collecting a subset of information from the source PDF document.
    The process extracts the title, keywords, and abstract from the header; and paragraphs, sections. and figure and table captions from the body.
    All the callouts to references, tables, and figures are ignored.
    The resulting structured document is then encoded in XML as will be described below. 
    \item XML to the annotation tool exchange formats: We transform our XML-formatted data into an INCEpTIONS compatible import format, such as the Webanno TSV 3.2 (\url{https://inception-project.github.io/releases/0.17.0/docs/user-guide.html\#sect_formats_webannotsv3}), and vice-versa using a set of Python scripts. 
    The Webanno TSV 3.2 format is an extension of the CONLL (\url{https://www.signll.org/conll/}) format, with additions of the header and column representation.
\end{itemize}

\section{Data Record}
\label{sec:data-record}
\begin{figure}[ht]
\centering
\subfloat[Papers distribution by Licence: Open Access vs copyrighted.]{\resizebox*{4cm}{!}{\includegraphics{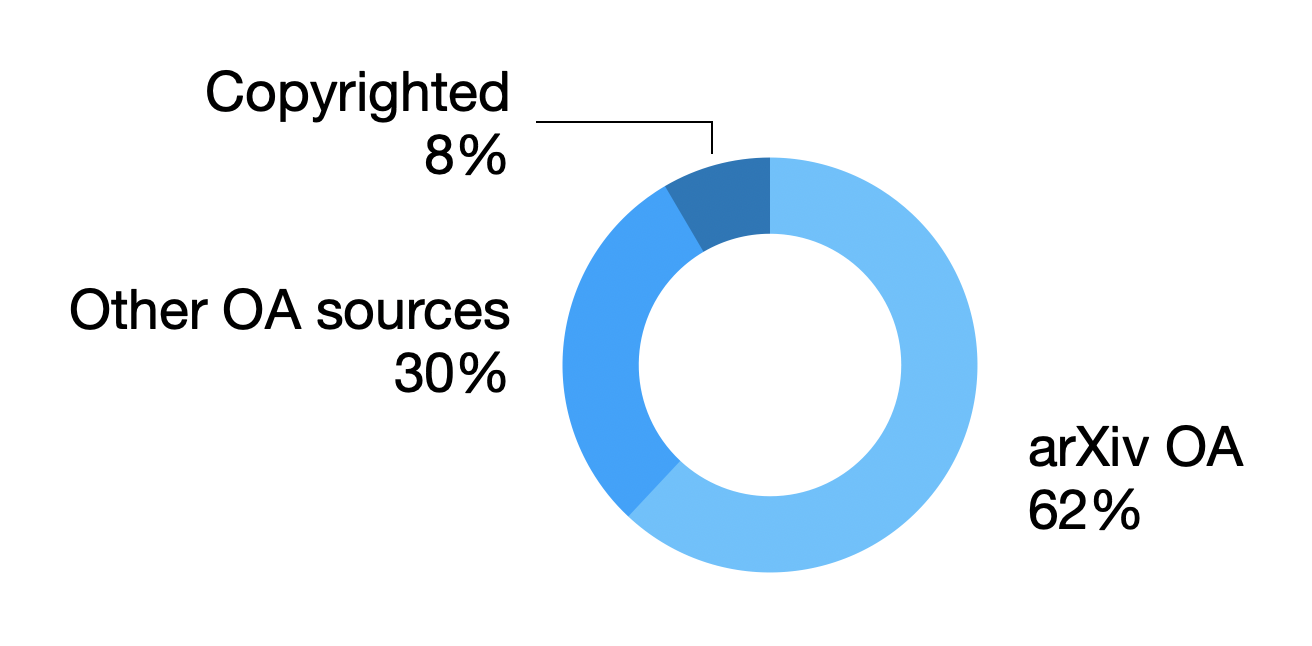}}\label{fig:arxiv-rate}}
\hspace{5pt} 
\subfloat[Distribution by publisher.]{\resizebox*{4cm}{!}{\includegraphics{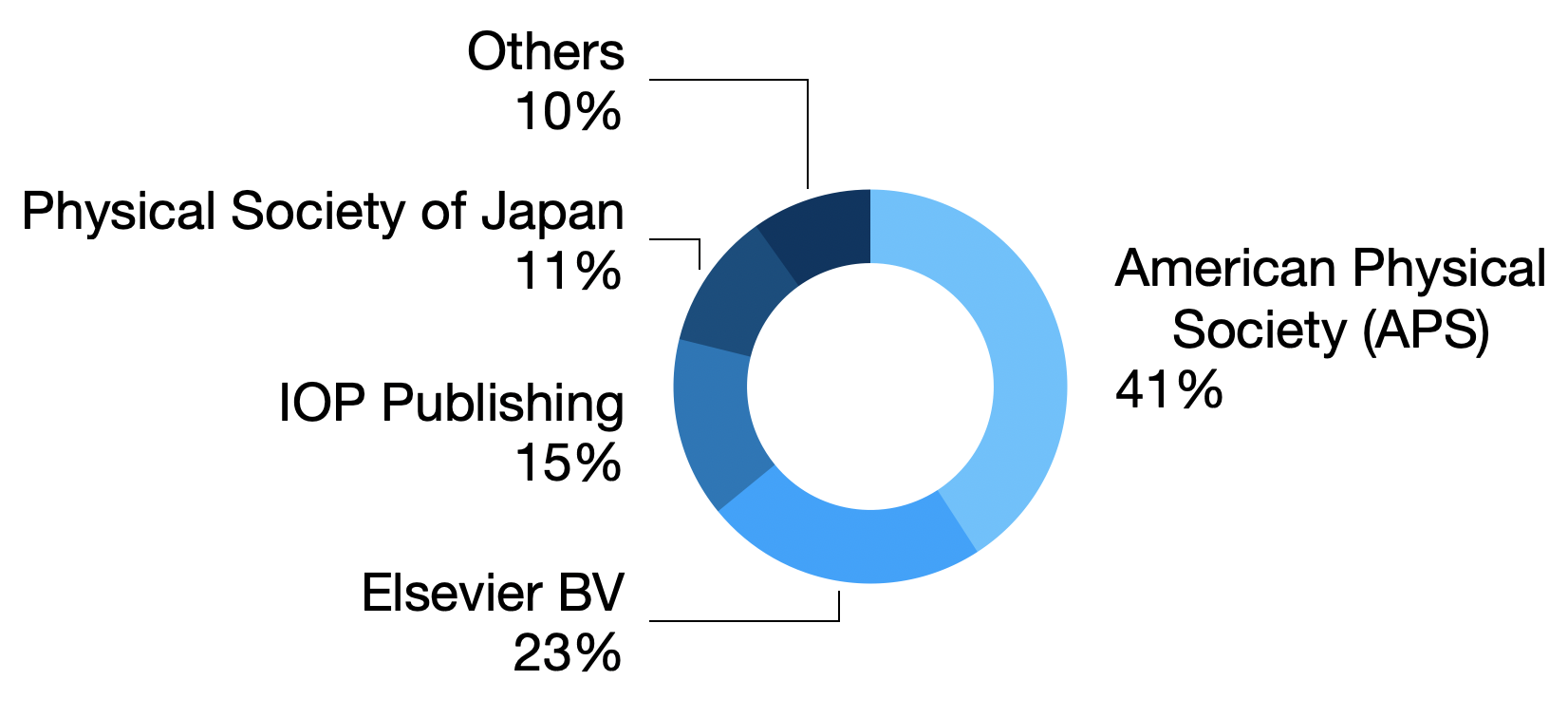}}\label{fig:distribution-by-publisher}} 
\hspace{5pt} 
\subfloat[Distribution by year of publication.]{\resizebox*{4cm}{!}{\includegraphics{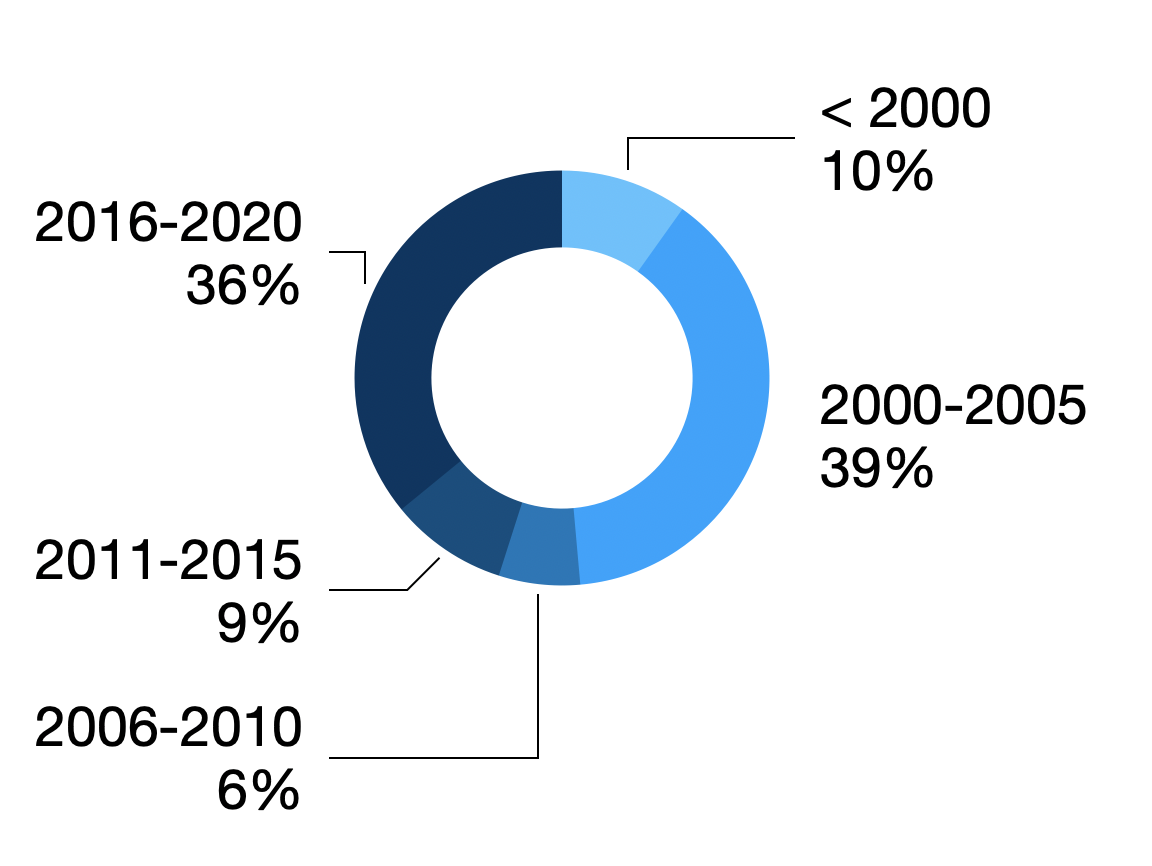}}\label{fig:distribution-by-year}} 
\caption{Distribution of paper in the dataset by (a) license, (b) publisher, and (c) year of publication.} 
\label{fig:dataset-distributions}
\end{figure}

The dataset is composed of 142 PDF documents, of which 92\% (130) are OA (Figure~\ref{fig:arxiv-rate}).
To comply with copyright restriction, few articles from our dataset are not publicly available in our repository. 
The top three publishers represented in the corpus are American Physical Society (APS), Elsevier, and IOP Publishing (Figure~\ref{fig:distribution-by-publisher}).
Figure~\ref{fig:distribution-by-year} illustrate the distribution by publication date.
We summarise SuperMat's content in Table~\ref{table:summary-content}, with the statistics of documents, entities, and links given separately. In particular, this dataset contains 16052 (7166 unique) entities spread over six labels and 1398 links. 

Each document is encoded according to the XML TEI guidelines, which is a rich format for document representation. 
We have carried out no specific customisation, in order to remain fully compliant with the general TEI schema.
A TEI document has two main parts: the header (within the \texttt{<teiHeader>} tags) containing all the document metadata, and the body (within the section delimited by the \texttt{<text>} tag). 
The transformed data has the following structure: 

\begin{verbatim}
<TEI xml:lang="en" xmlns="http://www.tei-c.org/ns/1.0">
    <teiHeader>
        <fileDesc>
            <titleStmt>
                <title>[...]</title>
            </titleStmt>
            <publicationStmt>
                <publisher>[...]</publisher>
            </publicationStmt>
        </fileDesc>
        <encodingDesc/>
            <abstract>
                <p>[...]</p>
                <ab type="keywords">[...]</ab>
            </abstract>
        <profileDesc>
        </profileDesc>
    </teiHeader>
    <text>
        <body>
            <p>[...]</p>
            <ab type="tableCaption"> [...] </ab>
            <p> [...] </p>
            <ab type="figureCaption"> [...] </ab> 
        </body>
    </text>
</TEI>
\end{verbatim}

We transformed the source documents into these TEI-compliant structures using a simplified representation for specific content types.
The general objective is to flatten the content into a generic structure where priority is given to the annotations.
For instance, the keywords section, which groups together the key terms defined by the author(s) of the paper, is encoded using the generic tag \texttt{<ab type="keywords">} as free text, instead of the dedicated \texttt{<keywords>} element that would typically be part of the header. 
For both the abstract and the article body, the text is segmented in paragraphs (by means of the \texttt{<p>} element). 
The text is annotated with the generic \texttt{<rs>} (referencing string) element adorned with three attributes: \texttt{@type} (the entity type), \texttt{@corresp} (to provide a link to another annotation such as from \textit{material} to \textit{T\textsubscript{c}}), and \texttt{@xml:id} (to uniquely identify the annotation for referencing or linking purposes).

Because only the captions of tables and figures are retained from the original source, a simplified encoding was defined by means of the \texttt{<ab>} element characterised by a \texttt{@type} attribute; that is, \texttt{<ab type="figureCaption">} for figure captions and \texttt{<ab type="tableCaption">} for table captions. 
Here is an example: 

\begin{verbatim}
<p>
    The electron-doped high-<rs type="tc">transition-
    temperature</rs> (<rs type="tc">Tc</rs>) <rs 
    type="class">iron-based pnictide</rs> 
    superconductor <rs type="material" 
    xml:id="m6">LaFeAsO1-xHx</rs> has a unique 
    phase diagram: Superconducting (SC) double domes are 
    sandwiched by antiferromagnetic phases at ambient 
    pressure and they turn into a single dome with 
    a maximum <rs type="tc">Tc</rs> that 
    <rs type="tcValue" xml:id="m7" 
    corresp="#m6,#9">exceeds 45K</rs> 
    at a pressure of <rs type="pressure" 
    corresp="#m7">3.0 GPa</rs>. 
    [...]
</p>
\end{verbatim}

In the above snippet, the entities \textit{"3.0 GPa"}, \textit{"exceed 45K"} and \textit{"LaFeAsO1-xHx"} are linked together via the pairs \texttt{@corresp, @xml:id}. 
This schema supports multiple annotations to any part of the document. 
For example, the entity \textit{exceed 45K} has a second link with the corresponding identifier (\textit{"\#9"}) to an annotation outside this paragraph.

\section{Applications}
\label{sec:applications}
SuperMat is constructed as a resource for TDM applications in superconducting materials. It can be used as data source in several complementary tasks: 
(1) creation of an automatic information extraction system for dataset creation,
(2) articles classification, 
(3) named entity extraction (for example, automatic dictionary construction), 
(4) clustering and document synthesis,
(5) training of machine learning (ML) algorithms,
(6) evaluation of rule-based or ML-based algorithms, and 
(7) development of downstream processes, such as material name parser, or quantity normalisation.

\subsection{Practical applications}
Such a dataset may benefit several types of possible applications: 

\begin{itemize}
    \item Evaluation tasks: This corpus can be used for evaluation tasks on automatic extraction. In particular, we can envision two popular tasks in superconducting materials science, namely: (a) NER and (b) EL methods. EL techniques have been mainly designed and studied using text from Wikipedia and newswires services which represent most of the available data. 
    To the best of our knowledge, however, there is no application within materials science.
    \item Automatic information extraction for superconducting materials: This dataset can be used as training data for such a purpose. 
    Automatic information extraction using ML and text mining techniques can accelerate the construction of databases for superconducting materials.
    \item Document retrieval: Information retrieval is a key application helping researchers overcome information overload.
    One way is through query expansion to cover multiple expressions of the same term. 
    By collecting and clustering all expressions under the same concept, it would be possible to retrieve documents when, for example, the resistivity measurement is described by a phrase other than "resistivity". 
    Furthermore, the assigned labels can be used to boost documents where a certain term belongs on a specific label. 
    For example, \textit{cobalt oxide} can appear as either \texttt{<material>} or \texttt{<class>} depending on the context, while a user would like to obtain documents where \textit{cobalt oxide} appears as \texttt{<material>}.
    \item Weighted-clustering: Scientific document clustering has recently gained growing attention because of its potential capacity for finding additional relevant documents of interest.
    For example, clustering can help locating similar experimental settings in a large collection of documents. However, clustering documents based on their general content might not be optimal for finding such detailed similarities.
    Annotation can be leveraged to tilt the clustering algorithm toward entity similarity, which may provide a more focused clustering towards a specific type of information.
\end{itemize}

\label{sec:technical-validation}
\section{Technical Validation} 
The following measures were employed to ensure the creation of a high-quality dataset: 
\begin{itemize}
    \item Each document was revised and validated by domain experts, 
    \item The workflow begins by assigning machine-based annotated data. This has demonstrated to improve the annotation task over several aspects, namely: time consumption, error rate, and annotation agreement~\cite{Fort2010InfluenceOP,Nvol2011SemiautomaticSA,Lingren2014EvaluatingTI}.
    \item On-the-fly automatic annotation recommendations, which provide fresh suggestions based on online decisions made by the annotators.
    \item The annotators have rapid access to changes in the annotation guidelines.
    \item The discussions were documented and linked in the guidelines. 
    \item Reviews are discussed and approved collaboratively between domain experts and other annotators.
\end{itemize}

These guidelines are a vital piece of this work since they contain knowledge accumulated from these activities.
However, measuring the completeness of the guidelines is challenging. 
Assuming that the documents validated by domain experts represent the ground truth, we conducted IAA analysis between different annotators against the ground truth, using the Krippendorf's Alpha metric~\cite{Krippendorff2004ReliabilityIC}.
Table~\ref{table:average-iaa} shows the average IAA which is satisfying with a value of approximately 0.9. 
The highest score is obtained in the \texttt{<material>} entities, while the lowest one is obtained in \texttt{<pressure>}, which appears less frequently in the papers. 
The disagreement in \texttt{<tcValue>} can appear to be too low as compared with other labels such as \texttt{<class>}, which is, at first look, more ambiguous. 
We analysed the different cases and identified three reasons why this happens. 
First, \texttt{<tcValue>} may depend heavily on the context that requires more human attention, and it is therefore more prone to errors. 
Second, our suggestions system is challenged in its ability to disambiguate critical temperatures from other temperature data, leading to incorrect or invalid suggestions. 
Finally, the presence of mathematical symbols (e.g. "\texttt{\~}", "\texttt{<}", and "\texttt{>}") or other modifiers ("\texttt{up to}", "\texttt{exceeds}", etc.) before the \texttt{<tcValue>} could generate small disagreements that accumulate in the average score. 

To more precisely isolate the impact of the guidelines, we grouped the IAA results by level of domain experience. 
Table~\ref{table:comparison-iaa-nde-de} displays the IAA between the validated data and the data corrected by (a) domain experts (researchers who conduct superconducting development experiments), (b) non-domain-experts (researchers with no experience with superconducting materials), and (c) novices (students in materials science with limited domain experience). 
Obviously, the domain experts have the highest agreement and the IAA value (around 0.95) is 0.06 higher on average than that of non-domain experts. 
Thus, superconducting materials is a complex domain that requires knowledge in materials science to produce high-quality data, while crowdsourcing initiatives such as the Amazon Mechanical Turk might not work well. 

Furthermore, we measured the reliability of the guidelines by observing how quickly novices could reach a satisfying agreement with the validation of the domain experts, without any previous training on the guidelines.
From Table~\ref{table:comparison-iaa-nde-de}, the  novices can attain high IAA results by only using the guidelines and our annotation support tools. 
The average difference in agreement with domain experts (around 0.05) indicates that the guidelines are precise and complete, and that the annotations tools offer sufficient support. 

\section{Data Availability}
\label{sec:code-availability}
The references of the original papers and their bibliographical information, the annotation guidelines sources, and the developed code are available at the GitHub repository \url{https://github.com/lfoppiano/SuperMat}. The guidelines are freely accessible at \url{https://supermat.readthedocs.io}.
The data transformation scripts were written in Python and can be run from the command line. 
They require BeautifulSoup  (\url{https://www.crummy.com/software/BeautifulSoup/}), an open-source library for parsing XML and HTML formats. 
The data analysis scripts were developed as Jupyter notebooks (\url{https://jupyter.org/}) which can easily output results and graphs in the browser. 
The open source annotation tool is INCEpTION (\url{https://inception-project.github.io/}). 
The content was acquired using biblio-glutton (\url{https://www.github.com/kermitt2/biblio-glutton}) and Grobid (\url{https://www.github.com/kermitt2/grobid}).
We computed the IAA using the Java library DkPro statistics (\url{https://dkpro.github.io/dkpro-statistics/})~\cite{Meyer2014DKProAA}.

\section{Conclusions}

In this paper we described the construction of an annotated linked dataset from scientific publications on superconductors development. 
SuperMat aims to establish a solid infrastructure where to build or improve TDM processes in superconductor materials domain.
We annotated 142 full-text articles where the data was automatically extracted from the PDF document and encoded through the XML TEI guidelines providing a basic structure of the original document. 
The dataset is validated by domain experts and provides 16052 entities of six categories, and 1398 links between materials, properties and conditions. 
This approach can be extended to other materials domains following similar methodology.

\section*{Acknowledgements}

We would like to thank Tanifuji Mikiko for her continuous support, as well as the enthusiasm and the openness with which she lead the Data PlatForm Data Center (DPFC, \url{https://www.nims.go.jp/eng/research/materials-data-pf/index.html}) at NIMS.
Our warmest thanks to Patrice Lopez, the author of Grobid (\url{https://github.com/kermitt2/grobid}) and other TDM open-source projects. 

\section*{Notes on contributors}

L.F. designed and developed the work (data preparation, annotation tools, IAA experiments, automatic annotations). M.I. and Y.T. supervised the project.
L.R. defined the standardised dataset TEI format.
L.F. S.D. A.S. A.U. M.G.E.E. P.B.C. Y.M. S.I., and K.T. performed the dataset annotation and validation.
M.G.E.E. P.B.C. Y.M. S.I. K.T., and Y.T. validated the corpus.
L.F. wrote the manuscript with assistance in editing from S.D., M.I., K.T., P.B.C., Y.M., and  M.G.E.E.. 
All authors reviewed and approved the final manuscript. 

\section*{Competing interests} 

The authors declare no competing interests.

\bibliographystyle{tfnlm}
\bibliography{references}  

\section*{Figures \& Tables}

\begin{table}[ht]
    \tbl{Summary of the IAA for each annotation iteration.}
    {\begin{tabular}{ ccc } 
    \toprule
        \textbf{Iteration} \# & \textbf{IAA} & \textbf{IAA by label}  \\ [0.5ex] 
    \midrule
        1  & 0.45
        &\begin{tabular}{  cc  } 
            \texttt{<material>} & 0.45\\ 
            \texttt{<tc>} & 0.56\\
            \texttt{<tcValue>} & 0.50\\
            \texttt{<doping>} & 0.21\\
        \end{tabular}    
        \\ 
    \midrule
        2 & 0.65
        &\begin{tabular}{  cc  } 
            \texttt{<material>} & 0.75\\ 
            \texttt{<tc>} & 0.85\\
            \texttt{<tcValue>} & 0.85\\
            \texttt{<doping>} & 0.39 \\
        \end{tabular}          
        \\ 
    \midrule
        3 & 0.89
        & \begin{tabular}{  cc  } 
            \texttt{<material>} & 0.89\\ 
            \texttt{<tc>} & 0.91\\
            \texttt{<tcValue>} & 0.88\\
            \texttt{<doping>} & 0.94\\
        \end{tabular}       
        \\ 
    \bottomrule
    \end{tabular}}
    
    \label{table:summary-preliminary-annotation}
\end{table}

\begin{table}[ht]
     \tbl{Inconsistencies resulting from the overlapping of \texttt{<material>} and \texttt{<class>} labels.}
     {
    \begin{tabular}{ ccccc } 
    \toprule
        Text & Label 1 & \# & Label 2 & \#\\
    \midrule
        \texttt{superconducting transition}     &   \texttt{<material>}   &    1   &   \texttt{<tc>}  &   61   \\
        \texttt{NCCO}    &	\texttt{<material>}   &    14   &   \texttt{<tc>}  &   1   \\
        \texttt{superconducting transition temperatures}     &   \texttt{<material>}   &    1   &   \texttt{<tc>}  &   11   \\
        \texttt{occurrence of superconductivity}    &	\texttt{<material>}   &    1   &   \texttt{<tc>}  &   1   \\
    \bottomrule
    \end{tabular}}
    \label{table:dataset-inconsistencies-unclear}
\end{table}

\begin{table}[ht]
    \tbl{Inconsistencies resulting from human mistakes.}
    {\begin{tabular}{ ccccc } 
    \toprule
        Text & Label 1 & \# & Label 2 & \#\\
    \midrule
        \texttt{LiFeAs}         &   \texttt{<material>}   &    89   &   \texttt{<class>}  &   1   \\
        \texttt{Bi-2212}        &	\texttt{<material>}   &    34   &   \texttt{<class>}  &   1   \\
        \texttt{cobalt oxide}   &   \texttt{<material>}   &    89   &   \texttt{<class>}  &   1   \\
        \texttt{RE-123}         &	\texttt{<material>}   &    34   &   \texttt{<class>}  &   1   \\
    \bottomrule
    \end{tabular}}
    \label{table:dataset-inconsistencies-clear}
\end{table}

\begin{table}[ht]
    \tbl{Statistical overview of the dataset. 
    Links\textsubscript{ip} indicates the number of links within the same paragraph (intra-paragraph). Links\textsubscript{ep} indicate the number of links from different paragraphs  (extra-paragraphs).  }
    {\begin{tabular}{ m{6em}   m{4em}  m{6em}  m{7em}  m{6em} } 
    \toprule
        \multirow{2}{5em}{\textbf{Documents}} & \textbf{Files} & \textbf{Paragraphs} &	\textbf{Sentences} & \textbf{Tokens}\\
         & 142  &	2800 & 	18344 & 	1118432\\
    \midrule
        \multirow{2}{5em}{\textbf{Entities}} & \textbf{Entities} &  \multicolumn{2}{|c|}{\textbf{Unique entities}} &  \textbf{ Labels} \\
        & 16040 &  \multicolumn{2}{c}{7151} &  6 \\
    \midrule
        \multirow{2}{5em}{\textbf{Links}} & \textbf{Links} & \multicolumn{2}{|c|}{\textbf{Links\textsubscript{ip}}} 
        & \textbf{Links\textsubscript{ep}}\\
        & 1399  & \multicolumn{2}{c}{1286} &	113	\\
    \bottomrule
    \end{tabular}}
    \label{table:summary-content}
\end{table}

\begin{table}[ht]
    \tbl{Average IAA between the annotated and validated documents}
    {\begin{tabular}{ cc } 
    \toprule
        \textbf{Label} & \textbf{Average}\\
    \midrule
        \texttt{<material>}     &   0.956   \\
        \texttt{<me\_method>}   &	0.887   \\
        \texttt{<pressure>}     &	0.723   \\
        \texttt{<class>}        &	0.925   \\
        \texttt{<tcValue>}      &	0.863   \\
        \texttt{<tc>}           &	0.831   \\
    \midrule
        \textbf{Micro average}        &	0.911	\\
    \bottomrule
    \end{tabular}}
    
    \label{table:average-iaa}
\end{table}

\begin{table}[ht]
     \tbl{Calculated IAA for annotations produced by domain experts, non-domain experts, and novices compared to the validated version. Annotations from domain experts are cross validated. }
    {\begin{tabular}{ ccccc } 
    \toprule
        \textbf{Label} & \textbf{Domain experts} & \textbf{Non-domain experts} & \textbf{Novices}\\
    \midrule
        \texttt{<material>}     &   0.969   & 0.950    &   0.924   \\
        \texttt{<me\_method>}   &   0.890   & 0.862    &   0.901   \\
        \texttt{<pressure>}     &   0.836   & 0.741    &   0.746   \\
        \texttt{<class>}        &   0.990   & 0.836	   &   0.899   \\
        \texttt{<tcValue>}      &   0.895   & 0.734	   &   0.841   \\
        \texttt{<tc>}           &   0.874   & 0.776	   &   0.830   \\
    \midrule
        \textbf{All labels}        &	0.940   &   0.882	&      0.896   \\
    \midrule
        \textbf{\# paragraphs}  &   1066   &  1648	    &   325     \\
    \bottomrule
    \end{tabular}}
   
    \label{table:comparison-iaa-nde-de}
\end{table}

\end{document}